\documentclass[pra,showpacs,twocolumn,longbibliography,secnumarabic,nofootinbib]{revtex4-2}
\usepackage{graphicx}
\usepackage{amssymb}
\usepackage{float}
\usepackage[T1]{fontenc}
\usepackage{amsmath}
\begin{document}
\title
{Environmental-induced work extraction}
\author{Rasim Volga Ovali$^{\dagger,1}$}
\author{Shakir Ullah$^{\dagger,2}$}
\author{Mehmet G\"{u}nay$^{\dagger,3}$}
\author{Mehmet Emre Tasgin$^{2,*}$}
\affiliation{$^\dagger$ Contributed equally}
\affiliation{$^*$ correspondence: metasgin@hacettepe.edu.tr and metasgin@gmail.com}
\affiliation{ $^{1}$Department of Physics, Recep Tayyip Erdogan University, 53100, Rize, Turkey}
\affiliation{ $^{2}$Institute  of  Nuclear  Sciences, Hacettepe University, 06800 Ankara, Turkey}
\affiliation{ $^{3}$Department of Nanoscience and Nanotechnology, Faculty of Arts and Science, Mehmet Akif Ersoy University, 15030 Burdur, Turkey}
\begin{abstract}
A local measurement extracts work as a backaction, e.g., in a system of two entangled cavities: first cavity, $a$, comprises a piston and the measurement is carried out on the second cavity, $b$. When no one makes a measurement on the cavity $b$, i.e., it is simply placed in vacuum; environmental monitoring results in the coherent states as the einselected pointer states~(the measurement basis)~[PRL 70, 1187 (1993)]. This makes the measurement, that nature itself performs, a Gaussian one with a fixed strength $\lambda=1$. We show that this makes nature assign a \textit{fixed} amount of work to a particular entanglement degree $0\leq \xi(r) \leq 1$, i.e., $W=\xi(r)\times(\bar{n}\hbar\omega_a)$, nothing that the term in parenthesis is the entire thermal energy. Afterwards, we show that this phenomenon applies quite generally, i.e, not restricted to a two-cavities system. We also touch on the influence of inherited symmterization entanglement in this context.
We can arrive an additional phenomenon by  considering that work is simply the process of converting randomly moving microscopic ingredients~(vanishing mean-velocity) into a directional one, i.e, with a nonzero mean-velocity. We show that such a change in the character of the motion introduces curvature in spacetime according to general relativity. This phenomenon is the first demonstration of a quantitative relation between entanglement and curvature using solely the quantum optics arguments. 
\end{abstract}
\maketitle

Quantum technologies, which shape the new century, are empowered by quantum entanglement and nonclassical states~\cite{acin2018quantum}. Measurements below the standard quantum limit~\cite{LigoNaturePhot2013,pezze2018quantum}, quantum enhanced imaging~\cite{Lugiato_JOptB_2002,casacio2021quantum,pirandola2018advances}, quantum radars~\cite{maccone2020quantum}, quantum teleportation~(QT)~\cite{bennett1993teleporting}, quantum computation~\cite{wright2019benchmarking} and information processing~\cite{vedral2006introduction} are all enabled by entangled ---more generally nonclassical~\cite{dodonov2002nonclassical,ge2015conservation}--- states. 

Quantum entanglement also refashions our understanding on thermodynamics. It can be employed as a resource in quantum heat engines~\cite{bresque2021two,josefsson2020double,hewgill2018quantum,zhang2007four,rossnagel2014nanoscale,dillenschneider2009energetics,scully2003extracting} which makes them operate more efficiently compared to their classical counterparts~\cite{kieu2004second}. Measurements are carried out on an ancilla which is entangled with the engine involving in the work extraction process. Information on the engine's state, extracted from the particular outcome of the ancilla-measurement, reduces the entropy of the engine system. 
This way, an extra amount of work proportional to the reduction in the entropy can be extracted~\cite{maruyama2009colloquium,lloyd1997quantum}.

\begin{figure*}
	\centering
	\includegraphics[width= 0.6 \textwidth]{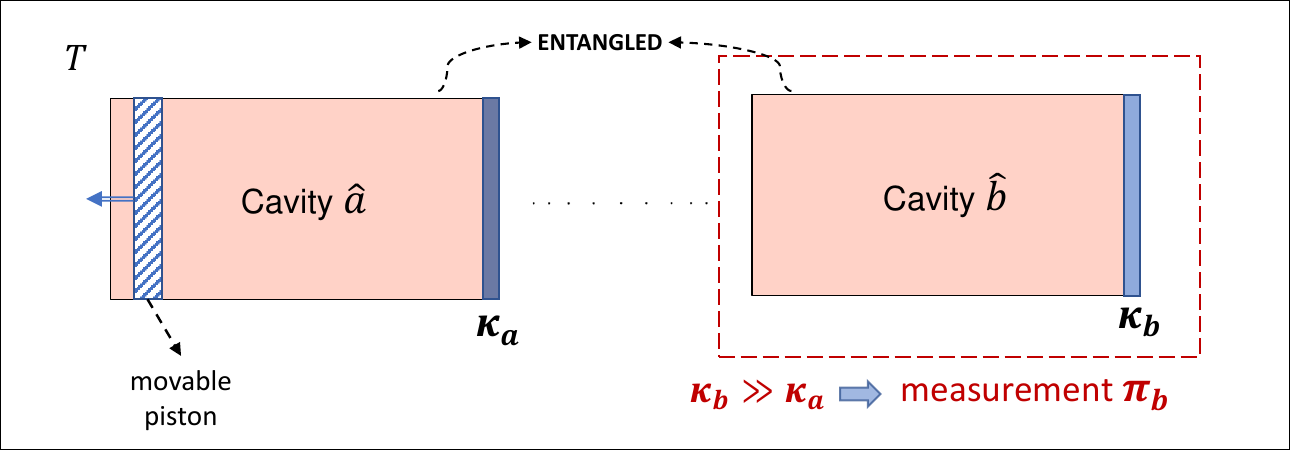}
	\newline
	\caption{Environmental-induced work extraction~(EIWE) in a two-entangled-cavities system. Cavity $a$, comprising a movable piston~\cite{brunelli2017detecting,cuzminschi2021extractable}, is relatively well isolated compared to the cavity $b$, i.e., $\kappa_a \ll \kappa_b$. Environment performs monitoring on the cavity $b$ (as $\kappa_b \gg \kappa_a$)~\cite{zurek1993preferred,schlosshauer2019quantum,zurek2003decoherence} and performs a Gaussian measurement with $\lambda=1$, because the measurement basis is coherent states~\cite{zurek1982environment,zurek1993coherent,gallis1996emergence,tegmark1994decoherence,wiseman1998maximally,zurek1994decoherence,zurek1995decoherence,paraoanu1999selection,zurek1993preferred}. This scheme will be extended to a more general one in Fig.~\ref{fig2} that is commonly encountered in nature.    }
	\label{fig1}
\end{figure*}

When it comes to continuous-variable photonic systems~\cite{braunstein2005quantum}, where entanglement is among electric/magnetic field components, Gaussian measurements~(GMs)~\cite{weedbrook2012gaussian} play the pivotal role. Gaussian states/measurements have some peculiar features. (i) A Gaussian state is completely~(uniquely) characterized by its covariance matrix $\sigma$. (ii) There is an even more intriguing feature for GMs. Let us imagine two coupled systems, e.g., $a$ and $b$ modes in two entangled cavities, e.g., Fig.~\ref{fig1}. Covariance matrix of the subsystem $a$ after performing a measurement $\hat{\Pi}_b$ on the $b$ subsystem, i.e.,  $\sigma_a^{\pi_b}$, is \textit{independent} of the measurement outcome in $b$~\cite{fiuravsek2007gaussian,weedbrook2012gaussian,fiuravsek2002gaussian,eisert2002distilling}. This result is quite important as both thermodynamical and quantum optics features of subsystem $a$ are solely determined by the covariance matrix $\sigma_a^{\pi_b}$~\cite{serafini2003symplectic,AdessoPRA2004}. This implies that entropy of the $a$-system $S_{\scriptscriptstyle V}(\sigma_a^{\pi_b})$ is independent of the measurement outcome in $b$~\cite{fiuravsek2007gaussian,weedbrook2012gaussian,fiuravsek2002gaussian,serafini2003symplectic,AdessoPRA2004}. It depends merely on the measurement-strength $\lambda$ as described below.

A general GM, i.e., any measurement involving auxiliary modes prepared in vacuum states employing both passive and active linear optics~(beam-splitters, phase-shifters and squeezers) and homodyne detection can be described by positive operator valued measure~(POVM)~\cite{fiuravsek2007gaussian,weedbrook2012gaussian,adesso2014continuous}
\begin{equation}
\hat{\Pi}_{b}(\alpha_{ b}) = \pi^{-1} \hat{D}_{b}(\alpha_{ b}) \: \gamma^{\pi_b} \: \hat{D}_{b}^\dagger(\alpha_{b}),
\label{PiB}
\end{equation}
where $\gamma^{\pi_b}=R(\phi) \: {\rm diag}(\lambda/2,\lambda^{-1}/2) \: R^T(\phi)$ is a single-mode covariance matrix issuing the strength of the measurement~($\lambda$) and $\hat{D}({\alpha_{b}})=\exp(\alpha_{b}\hat{a}_{b}^\dagger -\alpha_{b}^*\hat{a}_{b})$ is the displacement operator. Such measurements are not projective in general~\cite{weedbrook2012gaussian,adesso2014continuous}. $\lambda$ and $R(\phi)$ are~\footnote{\label{fn:Rphi} $R(\phi)$ stands for a rotation in the plane of $x$-$p$ quadratures. In our case, environmental-induced work extraction, it will not play any role as noise of a rotated coherent state is the same. This will become apparent in the following text.} determined only by the structure of the linear optical network~(measurement-basis). 
 Limits $\lambda\to 0$ and $\lambda\to \infty$, infinitely squeezed state in $x$ and $p$, correspond to homodyne detection of the $\hat{x}_{b}$ and $\hat{p}_b$ quadratures, respectively. $\alpha_{b}$ is the amplitude of the outcome for a particular measurement.

A measurement on the $b$ subsystem collapses the $a$ subsystem into the density matrix $\hat{\rho}_{{\scriptscriptstyle a}|\pi_b} = {\rm Tr}_b(\hat{\Pi}_b \: \hat{\rho}_{ab} \: \hat{\Pi}_b)/N$ with $N= {\rm Tr}(\hat{\Pi}_b \: \hat{\rho}_{ab} \: \hat{\Pi}_b)$~\cite{weedbrook2012gaussian,adesso2014continuous}. A general covariance matrix for a two-mode system can be expressed in the form~\cite{AdessoPRA2004,serafini2003symplectic}
\begin{equation}
\sigma_{ab}=
\begin{bmatrix}
	\sigma_a &  c_{ab}\\
	c_{ab}^T & \sigma_b 
\end{bmatrix}.
\label{sigma_ab}
\end{equation}
After the measurement on $b$, covariance of the $a$ mode reduces to~\cite{fiuravsek2007gaussian,weedbrook2012gaussian,fiuravsek2002gaussian,eisert2002distilling}
\begin{equation}
\sigma_a^{\pi_b} = \sigma_a - c_{ab}\: (\sigma_b + \gamma^{\pi_b})^{-1} \: c_{ab}^T 
\end{equation}
which depends only on the measurement strength ${}^{\ref{fn:Rphi}}$ $\lambda$ but not on the particular measurement outcome $\alpha_b$. We remind that $\gamma^{\pi_b}$, thus $\lambda$, depends only on the measurement setup. 

Therefore one arrives the following conclusion. If $\lambda$ is somehow fixed, e.g., for a fixed optical setup, a given amount of entanglement results a particular amount of reduction in the entropy $S_{\scriptscriptstyle \rm V}(\sigma_a^{\pi_b})$. The amount of extracted work can be expressed as the difference $W=k_B T\left[ S_{\scriptscriptstyle \rm V}(\sigma_a^{\rm eq}) -S_{\scriptscriptstyle \rm V}(\sigma_a^{\pi_b})  \right]$~\cite{maruyama2009colloquium,lloyd1997quantum}. Here $S_{\scriptscriptstyle \rm V}(\sigma_a^{\rm eq})$ is the von Neumann entropy of the $a$ mode when it is in equilibrium with its environment. The entropy of the engine~($a$-mode) before the $b$-measurement is $S_{\scriptscriptstyle \rm V}(\sigma_a^{\rm eq})$. After the measurement, entropy of the $a$-system drops to $S_{\scriptscriptstyle \rm V}(\sigma_a^{\pi_b})$ as a backaction of the measurement. When the engine rethermalizes with its environment its entropy increases again to $S_{\scriptscriptstyle \rm V}(\sigma_a^{\rm eq})$, but it extracts $W$ amount work in this process, e.g., by pushing the piston. 

Actually, there is nothing new up to here. That is, the above discussions/results have already been carried out in the literature~\cite{brunelli2017detecting,cuzminschi2021extractable}. Here, in difference, we examine the same phenomena within the concept of environmental-induced monitoring~\cite{zurek1982environment,zurek1993preferred,schlosshauer2019quantum,zurek2003decoherence} and demonstrate that quite interesting consequences show up. We consider two entangled cavities, supporting $a$ and $b$ cavity modes, one~($a$) is designed as an engine for extracting work and the second~($b$) is the one on which measurements are carried out, e.g., Ref.~\cite{brunelli2017detecting,cuzminschi2021extractable}, see Fig.~\ref{fig1}. We note that the two cavities do not need to be in direct contact in order to be prepared in an entangled state, i.e., swap mechanism. 

For the moment let us consider a relatively restricted setup, discuss the consequences on it and finally demonstrate that, actually, the phenomenon is commonly coincided in nature. 
We assume that the cavity $a$ is quite isolated from the environment with regards to the optical interactions, i.e., it is a high-quality cavity. Let one of the mirrors of the  cavity $b$ has relatively stronger interaction with the environment compared to the cavity $a$. The measurement-basis~(in other words the einselected pointer-basis~\cite{zurek1981pointer,zurek1982environment,zurek2003decoherence}) is determined by the characteristics of the measurement-setup. We raise the question: what is the measurement-basis if we simply leave the cavity $b$ in vacuum, i.e., if we do not place an optical measurement setup but leave the cavity $b$ alone?

The answer of this question, first given in 1993~\cite{zurek1993coherent} and studied so far extensively using various methods~\cite{gallis1996emergence,tegmark1994decoherence,wiseman1998maximally,zurek1994decoherence,zurek1995decoherence,paraoanu1999selection,zurek1993preferred}, is \textit{coherent-states}. The answer is the same if one considers, for instance, a harmonic chain~\cite{tegmark1994decoherence}. As the $a$-mode is well-isolated regarding the optical losses, we assume that environmental-induced monitoring~(measurement) takes place in the $b$-mode~\cite{zurek1993preferred,schlosshauer2019quantum,zurek2003decoherence}. (We will extend this setup to another one that is commonly encountered in nature. But at this stage this setup is a good choice for demonstrational purposes.) That is, when a cavity~(here $b$) is left in nature, the einselected pointer states ---the environmental monitoring ends up with~\cite{zurek1981pointer,zurek1982environment,zurek2003decoherence}--- are the coherent states. The measurement basis is coherent states~\cite{zurek1993coherent,gallis1996emergence,tegmark1994decoherence,wiseman1998maximally,zurek1994decoherence,zurek1995decoherence,paraoanu1999selection,zurek1993preferred}. This actually is a Gaussian measurement, Eq.~(\ref{PiB}), having the measurement-strength $\boldsymbol{\lambda=1}$ in the covariance matrix $\gamma^{\pi_b}$. Now, $\gamma^{\pi_b}$ becomes independent from the rotation $R(\phi)$, as a rotated coherent state is also a coherent state.

Therefore, the first main result of the paper, i.e., environmental-induced work extraction~(EIWE), can be stated as follows.

{\it Environmental monitoring~\cite{zurek1993preferred,schlosshauer2019quantum,zurek2003decoherence} performs a Gaussian measurement  ~\cite{zurek1993coherent,gallis1996emergence,fiuravsek2007gaussian,weedbrook2012gaussian,adesso2014continuous} on a cavity left in nature, i.e., without the influence of an intellectual being like a demon, human or an ancilla~\footnote{We also discuss what happens if there exists, e.g., some masses of atoms in surrounding in the framework of harmonic chains~\cite{tegmark1994decoherence}.}. The measurement-basis is the coherent states. As the measurement-strength is fixed $\lambda=1$, the cavity/engine~($a$), entangled with the measured mode $b$, extracts a particular amount of work $W$ for a given degree of entanglement $\xi(r)$. At low temperatures, e.g., room temperature for optical modes, extracted work is related with the entanglement as follows. (See supplementary material for the simple derivations.) }
\begin{equation}
W=\xi(r) \times \left(  \bar{n} \hbar \omega_a \right), 
\label{Wxir}
\end{equation}	
{\it where degree of the entanglement $\xi(r)=1-2/(1+\cosh2r)$ ranges between $0 \le \xi(r)\leq 1$ depending on the two-mode squeezing~(entanglement) parameter $r$~\cite{ScullyZubairyBook}. }

One may realize that $(\bar{n} \hbar \omega_a)$ is the entire thermodynamical energy present in the cavity $a$, the one extracting the work. Thus, entanglement converts thermodynamical energy (i.e., zero mean-velocity) to an ordered one where particles~(piston) moves along a given direction with nonzero mean velocity. Even this last statement by itself is not new. What is novel in this paper is the following. The environmental monitoring~(conducted by nature itself) performs this conversion and it always converts a fixed rate~$\xi(r)$ of its total thermodynamical energy~$\left(  \bar{n} \hbar \omega_a \right)$ into work for a particular amount of entanglement $\xi(r)$, given in Eq.~(\ref{Wxir}). 

Here, we focus on the work--entanglement correspondence for a two-mode entangled system. However, the phenomenon is not limited to such kind of systems. Before demonstrating that analogs of such a two-cavity system commonly show up in the nature, we would like to note that a similar phenomenon also appears, e.g., in the entanglement of atomic ensembles with off-resonant light~\cite{tasginEIWE_Ensembles,bao2020retrodiction,hammerer2010quantum}. This is because, total spin operators behave as continuous variable operators~\cite{bao2020retrodiction,hammerer2010quantum} in ensembles with large number of atoms. Gaussian measurements are studied in the ensembles~\cite{tasginEIWE_Ensembles,bao2020retrodiction,hammerer2010quantum,girvin2019modern}, too. 

\begin{figure*}
	\centering
	\includegraphics[width= 0.7 \textwidth]{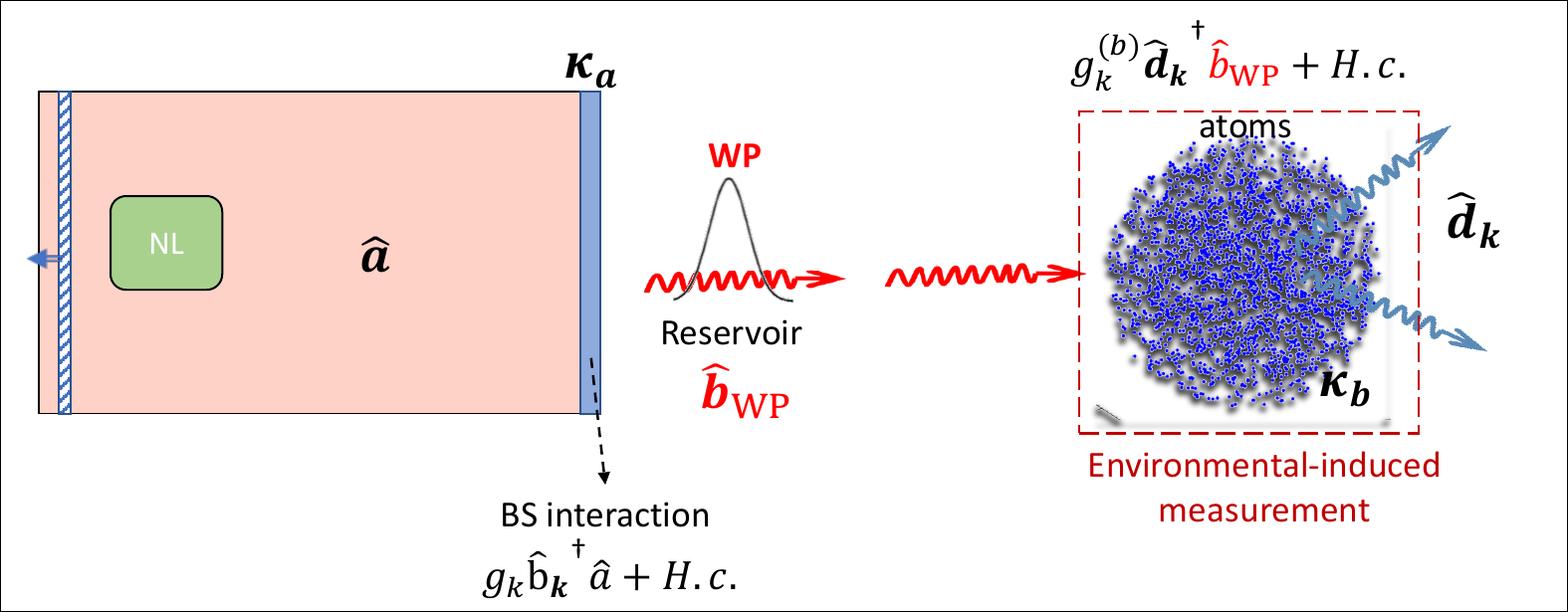}
	\newline
	\caption{Environmental-induced work extraction~(EIWE) is commonly encountered in nature. Off-resonant coupling to atoms/molecules is not different than a mirror regarding the environmental monitoring. Cavity $a$ contains nonclassical light due to nonlinearity~(green NL). Wavepacket~(WP) $\hat{b}_{\rm \scriptscriptstyle WP}=\sum_{\bf k} \beta_{\bf k} \hat{b}_{\bf k}$ leaking out from the cavity is entangled with $\hat{a}$. A measurement on  $\hat{b}_{\rm \scriptscriptstyle WP}$ is performed by the environmental modes $\hat{d}_{\bf k}$ via the mass of atoms~(blue dots). As the quality of the cavity $a$ is higher, i.e., $\kappa_a \ll \kappa_b$, environmental monitoring is conducted on the  $\hat{b}_{\rm \scriptscriptstyle WP}$ by $\hat{d}_{\bf k}$ modes.}
	\label{fig2}
\end{figure*}

{\bf \small Commonly encountered scheme}.--- We now demonstrate that work extraction via two-entangled-cavities system can be put into a more general form commonly encountered in the nature. Now, let us consider a cavity containing a nonclassical light. The nonclassicality may be generated by some nonlinear processes taking place inside the cavity.  The cavity mode~$\hat{a}$ leaks out of the cavity, see Fig.~\ref{fig2}. The leaked-out field~($\hat{b}_{\rm \scriptscriptstyle WP}=\sum_{\bf k} \beta_{\bf k} \hat{b}_{\bf k}  $~\cite{gardiner2004quantum}) is entangled with the cavity mode, because the cavity mirror generates a beam-splitter like interaction $\hbar(g_k \hat{b}_{\bf k}^\dagger\hat{a}+H.c.)$ between the two. Beam splitter interaction transforms the single-mode nonclassicality~(e.g., squeezing) into two-mode entanglement~\cite{PhysRevA.65.032323}. (Even a conservation-like relation holds between the two kinds of nonclassicalities~\cite{ge2015conservation}. This could make one consider squeezing as a potential work/ordering.) If the cavity contains a movable piston, measurement on the $\hat{b}_{\rm \scriptscriptstyle WP}$ mode extracts $W=\xi \times (\bar{n}\hbar\omega_a)$ amount of work from the cavity $a$. (WP stands for wave-packet~\cite{tasgin2020nonclassicality}.) Here $0\leq \xi \leq 1$ quantifies the degree of the entanglement between $\hat{a}$ and $\hat{b}_{\rm \scriptscriptstyle WP}$  similarly. 

Now, one needs to raise the question: in this scheme who performs the measurement on the $\hat{b}_{\rm \scriptscriptstyle WP}$? The answer of this question can be caught by reexamining the phenomenon `coupling of a cavity mode to outside'. Two light modes cannot interact directly. The coupling takes place over the mirror atoms. Coupling hamiltonian $\hat{\cal H}_{\rm int}=(\hbar g_k \hat{b}_{\bf k}\hat{a}+H.c.)$ is obtained by eliminating the atomic operators~\cite{chimczak2018creating,hetet2011single,reiter2012effective,chang2012cavity,xu2015input}. (Couplings of light to mirrors comprising of even several atoms are also studied in the recent decades~\cite{mirhosseini2019cavity,nie2023non}.)

The point we would like to arrive is that the leaked-out field $\hat{b}_{\rm \scriptscriptstyle WP}$ is monitored by the environment~$\hat{d}_{\bf k}$, see Fig.~\ref{fig2}, via its (off-resonant) coupling to the mass of atoms~\cite{chimczak2018creating,hetet2011single,reiter2012effective,chang2012cavity,xu2015input,mirhosseini2019cavity,nie2023non} present around. That is,  $\hat{b}_{\rm \scriptscriptstyle WP}$ itself is also coupled to the environment similar to the input-output formalism~\cite{gardiner2004quantum} or a series of them like in harmonic chains~\cite{tegmark1994decoherence}. This coupling, again eliminating the atomic operators, can be put into form $\hbar(g_k^{\rm (2)}\hat{d}_{\bf k}^\dagger \hat{b}_{\rm \scriptscriptstyle WP}+H.c.)$.~\footnote{The environmental modes $\hat{d}_{\bf k}$ are different than the $\hat{b}_{\bf k}$ modes. The reason is similar to the one for input-output formalism for a two-sided-cavity where we describe the vacuum modes on the left and right sides using different modes~\cite{gardiner2004quantum}. Here,  $\hat{d}_{\bf k}$  designates the modes coupled with the $\hat{b}_{\rm \scriptscriptstyle WP}$ mode through the masses around, i.e., not the $\hat{a}$ cavity.}

 When the coupling of the $\hat{d}_{\bf k}$ to $\hat{b}_{\rm \scriptscriptstyle WP}$ is sufficiently larger than the coupling of the  $\hat{b}_{\rm \scriptscriptstyle WP}$ to $\hat{a}$, i.e., $g_k^{\rm (b)} \gg g_k$,~\footnote{Cavity decay rates $\kappa$ are related to the coupling constants as $\kappa=\pi D(\omega) g_k^2$ where $D(\omega)$ is the  density of states~\cite{ScullyZubairyBook}.} environment $\hat{d}_{\bf k}$ sets the einselected pointer states within decoherence time scale~\cite{zurek1993preferred,schlosshauer2019quantum,zurek2003decoherence}. They are coherent states and the measurement is a Gaussian one with $\lambda=1$. 

Therefore, we find that the two-cavities setup is actually a commonly encountered scheme in nature. (See also biological aspects~\cite{polka2016tunable,dorfman2013photosynthetic,tamulis2014quantum,gauger2011sustained,o2018schrodinger}.) Thus, work--entanglement correspondence, set by environmental-monitoring, is a more general relation ---also including the light-ensemble entanglement schemes~\cite{tasginEIWE_Ensembles,bao2020retrodiction,hammerer2010quantum}. As nature itself performs the work extraction and changes the curvature~(see below), this stands rather like a law.

{\bf \small EIWE vs other measurements.}---  A maximum entangled state, e.g., $|e\rangle=(|1,0\rangle+|0,1\rangle)/\sqrt{2}$, also results in the extraction of the entire thermal energy similar to the $\xi=1$ case of EIWE given in Eq.~(\ref{Wxir}). Density matrix of such a system can be written as  $\hat{\rho}=P(|0,0\rangle\langle0,0| + x |e\rangle\langle e|)$~\footnote{We ignore other excited states, such as $|0,1\rangle$ and $|1,0\rangle$, which actually have the same energy. We do this for the sake of a reasonable comparison only.} in its thermal equilibrium, where $x=e^{-\hbar\omega_a/k_BT} \simeq \bar{n}$ is the mean number of photons. When one measures the number of photons  in the $b$-mode and outcome is $|1\rangle_b$, $W=x\hbar\omega_a=\bar{n} \hbar\omega_a$ amount of work is extracted in the cavity $a$. This is equal to the $\xi=1$ case of EIWE. However, one should note two differences. First, in this example, extraction of the same work is subject to the measurement of the $|1\rangle_b$ state unlike EIWE. Second, though both $|e\rangle$ and $\xi=1$ are max entangled states, the $\xi=1$ limit can be achieved only by keeping the two-mode-entanglement hamiltonian $(\hat{a}\hat{b}+H.c.)$ infinitely long times~\cite{ScullyZubairyBook}. The result is the same also for inherited symmetrization~(max) entanglement~\cite{tasgin2023energy}.

{\bf \small Symmetrization entanglement.}--- In our universe, there are collection of particles moving together~(i.e., into a single direction) as well as particles displaying randomized~(i.e., thermal) motion with respect to each other. From quantum thermodynamics we know that useful work~(single direction motion) is extracted at the expense of entanglement~\cite{bresque2021two,josefsson2020double,hewgill2018quantum,zhang2007four,rossnagel2014nanoscale,dillenschneider2009energetics,scully2003extracting}. In this paper, above, we also demonstrated that such a correspondence is already fixed [Eq.~(\ref{Wxir})] by nature itself. We now question if there could be a macroscopic source for the observed ordered motion in the universe~\cite{layzer1991cosmogenesis}. 

In Ref.~\cite{tasgin2023energy} we investigate the work corresponding to the symmterization entanglement. We examine the extracted work in the expense of breaking out a single particle's symmetrization-entanglement with the remaining condensate~\cite{tasgin2017many}. (We note that measurement and recoil in an interacting condensate are not a straightforward processes due to the collective behavior.) An expression similar to the max entanglement, $\xi=1$ in Eq.~(\ref{Wxir}), appears.  In brief, inherited sysmmetrization entanglement can be used to introduce an ordered motion, i.e.,  movement of particle(s) in a single direction. This phenomenon leads to interesting consequences, below.

{\bf \small Entanglement and curvature}.--- The potential for a correspondence between entanglement and curvature in spacetime intrigues the scientific community tremendously. Recent studies already presented connections between strong curvatures, i.e., wormholes, and entangled pairs~(e.g., electron-positron~\cite{sonner2013holographic,jensen2013holographic,kain2023probing}). Such a relation is first studied for two entangled black holes connected by a wormhole~\cite{maldacena2013cool}. These studies employ the holographic correspondence between 4-dimensional conformal field theory and AdS${}_5$ spaces. Another interesting study showed that entangled and not-entangled forms of a density matrix possess different weights~\cite{bruschi2016weight}: quantum field theory~(QFT) in a weakly curved spacetime is employed. These studies involve high energy physics toolboxes pronounced above.

Here, the result~(\ref{Wxir}) can be used to obtain a  direct relation between entanglement and curvature by employing quantum optics and general relativistic~(classical) arguments separately. After the work extraction process, merely the character of the particles' motion changes. Assuming a perfect entanglement $\xi=1$, the isotropic gas (i) moving homogeneously in all directions~(before the measurement, $v_{\rm mean}=0$, entanglement is present) transforms into a gas (ii) moving along a particular direction~(after the measurement, $v_{\rm mean}\neq 0$, entanglement vanishes). In Sec.~2 of the SM~\cite{SM}, we show that for a perfect relativistic fluid~\cite{brown1993action,misner1973gravitation} the Ricci scalar curvature differs by $\Delta R=R^{(ii)}-R^{(i)}=\frac{32 G p_0}{c^4}$ between the two instances. The difference emerges because the relativistic pressure becomes zero in the instance (ii). Here, $G$ is the gravitational constant and $p_0$ is the pressure. 

For a partial entanglement, i.e., $0<\xi<1$, a $\xi$ portion of the constituent particles move in a particular direction and curvature difference becomes $\Delta R= \xi \times \frac{32 G p_0}{c^4}$. That is, introduced curvature is proportional to the degree of the initial entanglement which diminishes after the environmental monitoring. Again, curvature is introduced by nature itself without the intervene of an intellectual being which makes entanglement-curvature relation something like law.  

We would like to remark that in obtaining this result we merely used quantum thermodynamics, i.e., not QFT in curved space. We judged the difference of the curvatures for the two instances, (i) and (ii), using the bare general relativity.  This derivation, in a manner, how Heisenberg obtained the uncertainty principle~(here the induced curvature in analogy) using merely the $p=h/\lambda$ formula~(here quantum thermodynamics) and classical mechanics~(general relativity), i.e., without employing quantum mechanics~(here QFT in curved spacetime) which is developed afterwards.

Finally, we would like to raise three remarks. First, a relation between entanglement and work/order is an already known (and to be expected) phenomenon. However, here we set this relation to a fixed value. Second, the relation is set by nature itself, so it is rather like a law here. This is similar for the entanglement-curvature connection $\Delta R= \xi \times \frac{32 G p_0}{c^4}$.
Third, while environmental-induced monitoring performs a Gaussian measurement~($\lambda=1$), not all quantum states need to be Gaussian, i.e., Eq.~(\ref{sigma_ab}). Nevertheless, studies show that many-body systems ---initialized at non-Gaussian states--- are shown to relax into Gaussian states~\cite{schweigler2021decay,cramer2008exact}, beyond the fact that Gaussian states already appear in numerous many-body systems~\cite{girvin2019modern} naturally.

\bibliography{bibliography}

\end{document}